\documentclass[twocolumn,english,aps,floatfix,superscriptaddress,showpacs,amsfonts,amssymb,superscriptaddress]{revtex4}
\usepackage{color}

\usepackage{graphicx}
\usepackage{amsmath}
\usepackage{latexsym}
\usepackage{epstopdf}
\usepackage{amsmath}
\usepackage{amssymb,amsmath}
\usepackage{multirow}
\usepackage{mathtools}
\newcommand{\beq}{\begin{equation}}
\newcommand{\eeq}{\end{equation}}

\def\ket#1{|#1\rangle}

\begin{document}

\title{Universal behavior of the Shannon and R\'enyi mutual information of  quantum critical chains}
\author{F.~C.~Alcaraz }

\affiliation{ Instituto de F\'{\i}sica de S\~{a}o Carlos, Universidade de S\~{a}o Paulo, Caixa Postal 369, 13560-970, S\~{a}o Carlos, SP, Brazil}

\author{M.~A.~Rajabpour}
\affiliation{ Instituto de F\'{\i}sica de S\~{a}o Carlos, Universidade de S\~{a}o Paulo, Caixa Postal 369, 13560-970, S\~{a}o Carlos, SP, Brazil}

\date{\today{}}

\begin{abstract}
We study the Shannon and R\'enyi mutual information (MI) in the ground state (GS) 
of different critical quantum spin chains. Despite the apparent basis dependence of these quantities we show the existence of  some particular basis (we will call them conformal basis) 
whose  finite-size scaling function is related to  the central charge 
 $c$ of the underlying conformal field theory of the model. 
 In particular, we verified that  for  large index $n$, 
the MI of a subsystem of size $\ell$ in a periodic chain with $L$ sites 
behaves as   $\frac{c}{4}\frac{n}{n-1}\ln\Big{(}\frac{L}{\pi}\sin(\frac{\pi \ell}{L})\Big{)}$, when the ground-state wavefunction is expressed in these 
special conformal basis. This is in agreement with recent predictions. For  generic local basis we will show that, although in some cases $b_n\ln\Big{(}\frac{L}{\pi}\sin(\frac{\pi \ell}{L})\Big{)}$ is a good fit to our numerical data, in general there is no direct relation between $b_n$ and the central charge of the system. We will support our findings with detailed numerical calculations  for 
the transverse field Ising model, $Q=3,4$ quantum Potts chain, quantum Ashkin-Teller chain and the XXZ quantum chain. We will also present some additional 
results 
 of the Shannon mutual information ($n=1$), for the  parafermionic $Z_Q$ quantum chains with $Q=5,6,7$ and $8$.
\end{abstract}
\pacs{11.25.Hf, 03.67.Bg, 89.70.Cf, 75.10.Pq}
\maketitle

\section{Introduction}

Quantum entanglement measures have been frequently  used recently to detect quantum phase transition in many body quantum systems. Measures like von Neumann
and R\'enyi entanglement entropy, concurrence and quantum discord are among the most frequently used ones, see for example \cite{Amico 2008,Modi2012}. One of 
the important reasons for the success of these measures in detecting quantum phase transition and ultimately identifying the universality class of quantum 
critical behavior of the system is the
simplicity in their calculation by using numerical techniques such as the power 
method  and the density matrix renormalization group (DMRG) \cite{Schollwock}. Since at the critical point one can usually
describe the system with a conformal field theory (CFT) it is natural to look for observables that can be related to the important quantities in CFT. This
 program has been carried out in one dimension with significant detail by relating the von Neumann and R\'enyi entanglement entropy of a bipartite system to 
the central charge of the underlying CFT, see for example \cite{CC2009}. Although these quantities can be calculated relatively easily by numerical calculations
they have been out of reach from experimental point of views. Recently another measure, the Shannon entropy,  
which is
based on specific measurements in the system \cite{Stephan2009},
has been also introduced in   the context of quantum critical chains.

The Shannon entropy of the system ${\cal X}$ is defined as
\begin{equation} \label{Shannon}
Sh({\cal X}) =-\sum_x p_x \ln p_x,
\end{equation}
where   $p_x$ is the probability of finding the system  in a configuration $x$. These probabilities, in the case where  {$\cal A$} is a  subsystem  of a quantum
chain with wave function 
$\ket{\Psi_{{\cal A} \cup {\cal B}}} = \sum_{n,m} c_{n,m}\ket{\phi_{\cal A}^n} 
\otimes \ket{\phi_{\cal B}^m}$, are given by the marginal probabilities 
$p_{\ket{\phi_A^n}} = \sum_m |c_{n,m}|^2$  of the subsystem ${\cal A}$, where 
$\{\ket{\phi_{\cal A}^n}\}$ and $\{\ket{\phi_{\cal B}}^m\}$ 
are the vector basis in subspaces $\cal A$ and $\cal B$. In our study we will always take the whole system ${\cal X}=L$  which also indicates
the size of the system then the subsystems $\cal A$ and $\cal B$ will be 
denoted by $\ell$ and $L-\ell$, respectively. 
We will call the Shannon entropy of a subsystem of size $\ell$ as the reduced Shannon entropy $Sh(\ell)$\cite{foot2}. Notice  that the Shannon entropy is basis dependent in 
opposite to the von Neumann entanglement entropy that is a basis independent 
quantity. However as  we will 
see along this paper,  
it also contains   universal aspects in a specific sense that we will clarify later.

As we will see in the next sections the reduced Shannon entropy has an extensive part which 
is non-universal. In order to extract this non-universal harmless part  it is 
useful to define the so called  Shannon mutual information.
It is defined as
\begin{equation} \label{mutual information}
I(\ell,L)= Sh(\ell)+Sh(L-\ell)-Sh(L),
\end{equation} 
where as before $Sh(\ell)$ and $Sh(L-\ell)$ are the reduced Shannon entropies of 
the subsystems and $Sh(L)$ is the Shannon entropy of the whole system. 
The  Shannon mutual information has an information theoretic meaning. It  
 is one of the measures used to quantify the amount of information shared among  two subsystems. 
It tells us  how much information one can get about the subsystem $L-\ell$ by doing measurements in the subsystem $\ell$ and vice versa. This quantity has been calculated numerically
for the quantum Ising model in \cite{Um2012,Wai2013} and for many other critical quantum spin  chains in \cite{AR2013}. It is worth mentioning that in \cite{Wolf2008}
it was proved  that the Shannon mutual information of classical 
systems, like the entanglement entropy, should also follow the area law. Recently there has been also some developments in calculating
the shannon and R\'enyi entropy of two dimensional quantum critical systems \cite{Alet2013,Alet2014}.
Note that by changing $Sh(\ell)$ with the von Neumann entanglement entropy in 
 (\ref{mutual information})  one can define the von Neumann mutual information 
which is a different quantity from  the Shannon mutual information $I(\ell,L)$.
 For recent developments  in this direction see \cite{bernigau,eisler}.

One can also generalize the above definitions to the R\'enyi entropy as
\begin{equation} \label{Renyi}
Sh_n({\cal X}) =\frac{1}{1-n}\ln \sum_x p_x^n.
\end{equation}
The $n\to 1$ limit gives back the Shannon entropy. Similarly one can also generalize the Shannon mutual information by using the above definition. 
We consider in this paper the 
 simple naive definition:  
\begin{equation} \label{mutual information Renyi}
I_n(\ell,L)= Sh_n(\ell)+Sh_n(L-\ell)-Sh_n(L).
\end{equation} 

Differently  from the entanglement entropy the Shannon  and R\'enyi entropies are both basis dependent, 
however, as we will study in this paper 
in some particular basis these entropies show universal behavior at the critical point  that  can be 
connected with the underlying CFT governing the long-distance physics at the 
quantum critical point. It is worth mentioning that these entropies 
were first studied in the context of 
 Rokhsar-Kilvelson wave functions \cite{Rokhsar1988,Ardonne2004} for two dimensional quantum systems.
 \cite{Stephan2009,Fradkin2006,Hsu2009,Oshikawa2010}. Based on the transfer 
matrix approach one can
map the 1D quantum chain into a 2D classical model. From this classical model 
  we can define a Rokhsar-Kivelson
wave function. It is  the wave function  of
a two dimensional quantum system expressed on basis with one-to-one 
correspondence with the configurations of the 2D classical model and whose coefficients are the corresponding Boltzmann weights.  It is shown in \cite{Stephan2009} that the Shannon entropy of the periodic quantum spin chain is equal to the entanglement entropy
of the half of the cylinder in the 2D Rokhsar-Kivelson wave function.

In this paper we will study the  Shannon and R\'enyi  mutual 
information in different quantum critical spin chains such as Ising model, Q-state Potts model, Askin-Teller model and the XXZ quantum chain. 
 We will restrict ourselves to the case where the quantum chains are 
in the pure state formed by their GS. We will 
 also analyse, in all these critical quantum chains, the importance of the basis used to express the wave functions.
 We will clarify which are the basis that possibly can have  a direct connection to the central charge of the system. In the conclusions we will also present 
the results for the Shannon mutual information of the $Z_Q$-parafermionic quantum chains, with $Q=5,6,7$ and $8$.

\section{Mutual  information in quantum spin chains}

In this section we  study different aspects of the Shannon and  R\'enyi entropies in 
the transverse field Ising chain, three and four-state Potts model, 
the Ashkin-Teller model and the XXZ chain. 
As it was already discussed  in \cite{Stephan2013} we should expect  a significant difference between the first four 
cases and the  last one.
We will start by discussing  the known conjectures about different cases and then we will present our numerical results  and, based on them, some  conjectures. We 
will largely emphasize in this paper the important role played by  the basis  used to calculate the different kinds of entropies.
In our study we will always confine ourselves to critical chains.

\subsection{Mutual information in the transverse field Ising spin chain} 

The Hamiltonian of this model  is given by 
\begin{eqnarray}\label{Ising}
H=-\lambda \sum_{i=1}^{L} \sigma_i^z\sigma_{j+1}^z-\sum_{i=0}^{L}\sigma_i^x,
\end{eqnarray}
where ($\sigma_i^z,\sigma_i^x$) are spin-1/2 Pauli matrices localized at the 
sites $i=1,\ldots,L$.
 The system is critical at $\lambda=1$.
The Shannon entropy of the periodic  system at the critical point was studied numerically in \cite{Stephan2009}
and \cite{Stephan2010}. The  numerical results suggested the
following form for the R\'enyi entropy of the GS of the whole chain:
\begin{equation}\label{EE in RK smooth boundary 1}
Sh_n(L)=\mu_n L+\gamma_{n},
\end{equation} 
where  $\mu_n$ and $\gamma_{n}$ are  non-universal and universal constants, respectively.
The numerical results for the universal constant term $\gamma_n$ for the periodic chain  with  
ground state wavefunction expressed in the $\sigma^z$ basis   
 are \cite{Stephan2010}

\begin{equation}\label{EE in RK smooth boundary}
\gamma_n(\lambda=1)=\left\{
\begin{array}{c l}      
    0 , & n<1\\
        0.2543925(5), &n=1\\
        \ln 2, & n>1 .
\end{array}\right.
\end{equation}
The discontinuity with respect to $n$ means that the replica trick is probably 
not suitable to calculate the standard Shannon entropy from the R\'enyi ones. 
The very interesting fact is the constant value of $\gamma_n$ for $n>1$. This  indicates that it  can  probably be calculated by looking to the asymptotic 
behavior $n\to\infty$ of $Sh_n$ in the $\sigma^z$ basis. 
  This observation has very interesting consequences when one considers the 
reduced R\'enyi entropy for the transverse field Ising model. 
Due to the ferromagnetic nature of the quantum chain 
the configurations
with the highest probability\cite{high-prob} in the Ising model are the ones with all the spins up or spins 
down, so in principle when one considers the reduced R\'enyi entropy the most 
important configurations are those with all the spins in the subsystem 
are up or down. The corresponding probability $\mathcal{P}$ is usually called emptiness formation probability (EFP)
and it has been calculated for conformal field theories in \cite{Stephan2013} and references therein. Introducing the logarithmic emptiness formation probability (LEFP) 
as $\mathcal{E}=-\ln \mathcal{P}$ one can summarize the result for the periodic boundary condition as \cite{Stephan2013}
\begin{eqnarray}\label{LEFP periodic BC}
\mathcal{E}(\ell)=a \ell+\frac{c}{8}\ln\Big{(}\frac{L}{\pi}\sin(\frac{\pi \ell}{L})\Big{)}+...,
\end{eqnarray}
where here and hereafter we denote by $...$  the sub-leading terms. The idea behind this calculation is as follows: 
the configuration with all spins  up,  in the $\sigma^x$ basis, can be seen in the two dimensional classical Ising model
as a free boundary condition. This happens because the classical spins in the transfer matrix approach
actually correspond to the eigenstates of the matrix $\sigma^z$.
Considering a CFT with a free boundary condition on the slit one can extract the above formula for the LEFP in the $\sigma^x$
basis \cite{Stephan2013}. The crucial point is that the free boundary conditions in the euclidean approach is  a conformal boundary condition \cite{Cardy1989} and so one can
 use CFT techniques.
One can follow a similar argument in the $\sigma^z$ basis:  it is not difficult to show that  fixing the spins in the $\sigma^z$ basis is equivalent of  fixing  the spins
in the two dimensional classical counterpart. This boundary condition is also a conformal
boundary condition and by following the arguments in \cite{Stephan2013} one can get the same formula as equation (\ref{LEFP periodic BC}). 

\begin{figure}
\begin{center}
\includegraphics[clip,width=0.9\linewidth]{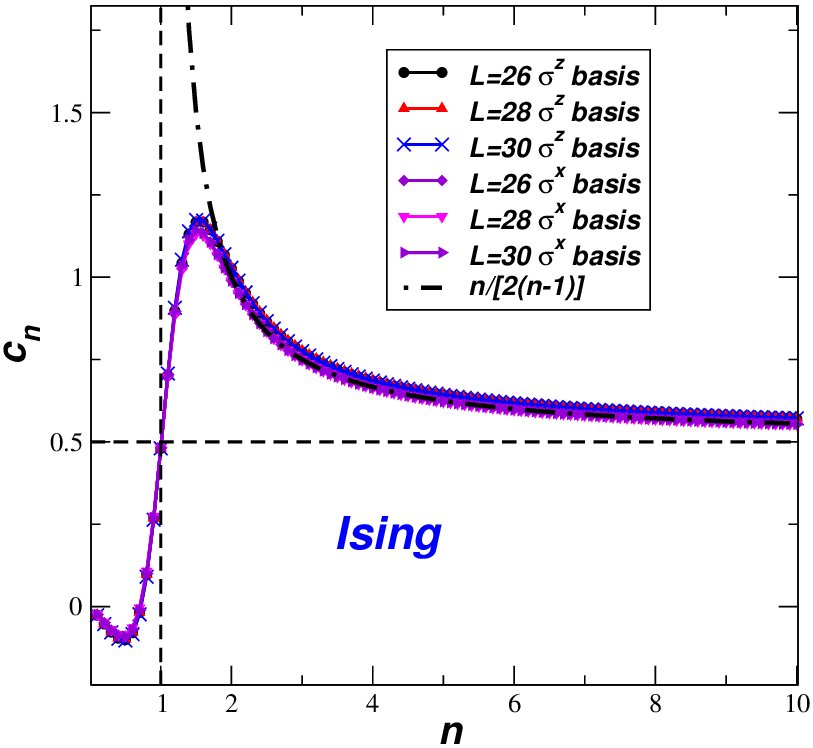}\\
\caption{\label{fig.1} (Color online)
Coefficient of the logarithmic term  of the R\'enyi MI in the Ising model in the $\sigma^z$ and $\sigma^x$ basis. The coefficients were found by restricting 
the fitting of (\ref{Mutual renyi information Ising}) to the subsystem sizes  $\ell=4,5,...,L/2$. 
The dashed straight lines are guidelines for $n=1$ and for the central charge $c=0.5$.}
\end{center}
\end{figure}

\begin{figure}
\begin{center}
\includegraphics[clip,width=0.9\linewidth]{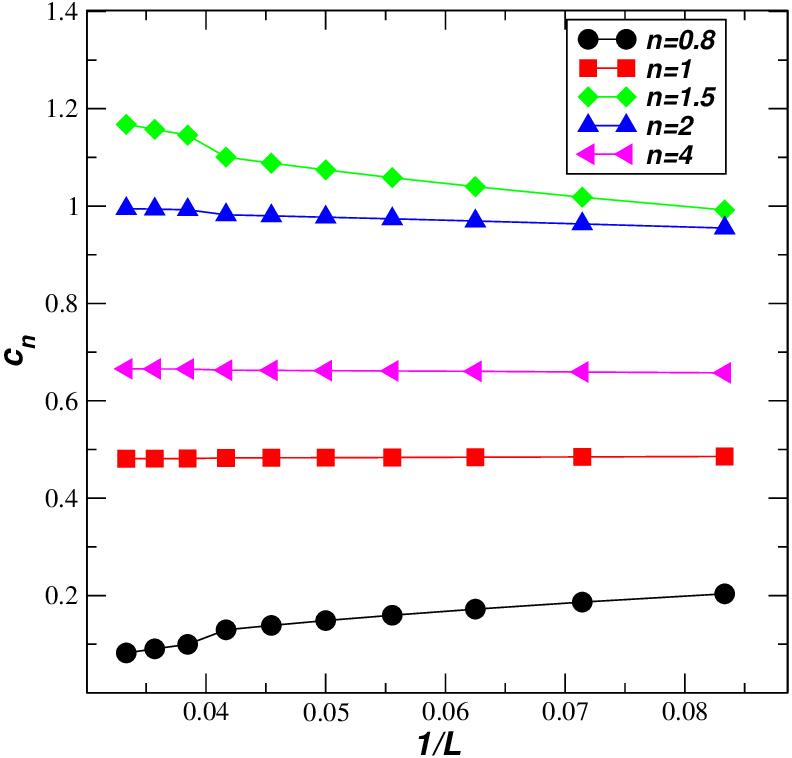}\\
\caption{\label{fig.2} (Color online)
Finite-size data of $c_n(L)$, for $L=12,14,\ldots,30$, for the GS Ising model in 
$\sigma^x$ basis. 
 The coefficients were calculated  by conditioning
the fitting to the subsystem sizes $\ell=4,5,...,L/2$.}
\end{center}
\end{figure}
\begin{figure}
\begin{center}
\includegraphics[clip,width=0.9\linewidth]{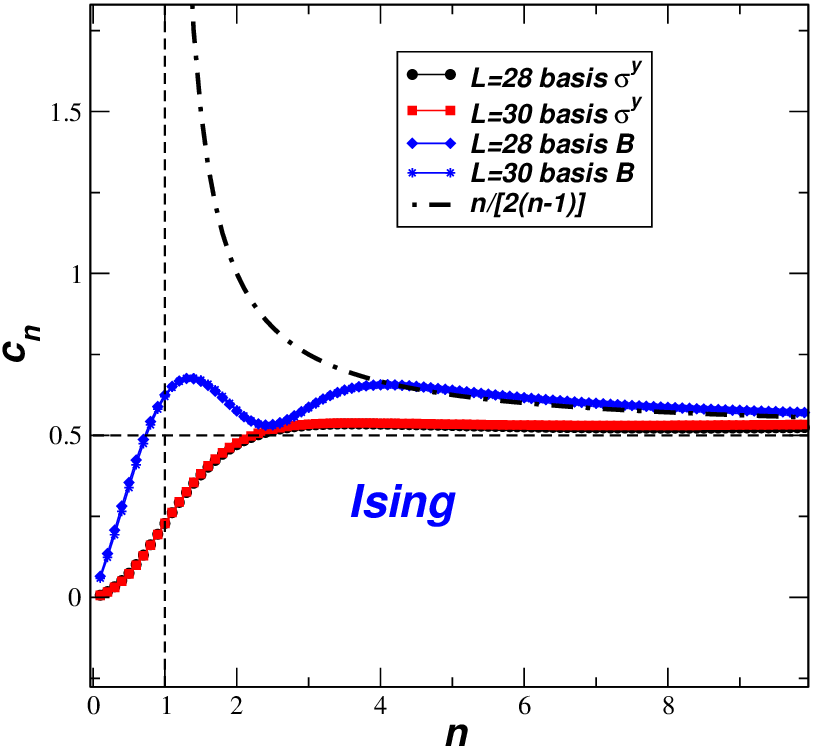}\\
\caption{\label{fig.3} (Color online)
Coefficient of the logarithmic term  of the R\'enyi MI in the Ising model in the $\sigma^y$ basis ($(\theta,\phi,\alpha)=(\frac{\pi}{4},0,0)$) 
and the $B$ basis ($(\theta,\phi,\alpha)=(\frac{\pi}{3},\pi,\frac{\pi}{5})$). The coefficients were found by conditioning
the fitting to the subsystem sizes $\ell=4,5,...,L/2$.
The dashed straight lines are guidelines for $n=1$ and for the central charge $c=0.5$.}
\end{center}
\end{figure}
 Using the LEFP and the fact that the behavior of the R\'enyi entropy for $n>1$ is controlled by $n\to \infty$ it was conjectured \cite{Stephan2013} that the 
reduced R\'enyi entropy of the GS 
should have the following form
\begin{eqnarray}\label{Renyi periodic BC Ising}
Sh_n(\ell)=\frac{n}{n-1}a \ell+\frac{c}{8}\frac{n}{n-1}\ln\Big{(}\frac{L}{\pi}\sin(\frac{\pi \ell}{L})\Big{)}+\gamma_n+...,
\end{eqnarray}
where $c=\frac{1}{2}$ is the central charge of the Ising model. As it was already 
mentioned one can not get the result for $n=1$ by  analytical continuation of the above result. 
Based on numerical results presented in our   previous work \cite{AR2013} we conjectured that the result for
$n=1$ is
\begin{eqnarray}\label{S periodic BC Ising}
Sh(\ell)=a \ell+\frac{c}{8}\ln\Big{(}\frac{L}{\pi}\sin(\frac{\pi \ell}{L})\Big{)}+\gamma_1+....
\end{eqnarray}

Based on the above formulas one can 
conjecture  the following formula for the  R\'enyi mutual information of
spin chains in the above two basis that are related to boundary CFT (from now on we will call them conformal basis) \cite{AR2013}
\begin{equation} \label{Mutual renyi information Ising}
I_n(\ell,L)= \frac{c_n} {4}\ln \left(\frac{L}{\pi}\sin(\frac{\pi\ell}{L})\right) + ...,
\end{equation}
where
\begin{equation}\label{c-n conformal Ising1}
c_n=c\left\{
\begin{array}{c l}      
    1, & n=1\\
\frac{n}{n-1}, & n>1 .
\end{array}\right.
\end{equation}

The above formula for $n=1$ has already been checked for many different quantum spin chains in \cite{AR2013} 
and the results looked consistent with the coefficient being very close to the central charge. However, 
recently \cite{Stephan2014} this result has been questioned in the case of Ising model,  where the numerical estimated value
 is $0.480$   instead of the
central charge value $c=\frac{1}{2}$.
In Fig.~1 we show the results of $c_n$ in the quantum Ising chain in the two different basis $\sigma^z$ and $\sigma^x$. 
These results were obtained by considering the fitting of 
(\ref{Mutual renyi information Ising}) considering the subsystem sizes 
$\ell=4,\ldots,L/2$.
The results confirm the validity of (\ref{c-n conformal Ising1}) 
 nicely for values of $n$ bigger than $n_c\sim 2$.
Taking spin chains with bigger lattice sizes might lead to a better compatibility with the formula (\ref{c-n conformal Ising1})
in the region $1<n<2$, see for example \cite{Stephan2014}. 
Our results also indicates  that the formula (\ref{Mutual renyi information Ising}) may also be valid for $0<n<1$ with the $c_n$ values   shown in
the Fig.~1 \cite{footnote}.

Let us make an important remark about the numerical results presented in Fig.~1, 
that will also be valid for all the subsequent numerical results presented in this 
paper. Although we obtained results for lattice sizes up to $L=30$ it is difficult 
to obtain reliable results for $c_n$ with precision smaller than a few percent 
by using extrapolating techniques. This is due to two reasons. The first one comes
 from the fact that the finite-size estimator $c_n(L)$, for a given lattice 
size $L$, is obtained from a fit of the data to (\ref{Mutual renyi information Ising}), in which the effect of a given sublattice 
size $\ell$ is distinct for each lattice size $L$. In Fig.~2 we show the finite 
estimators $c_n(L)$, for $L=12,14,\ldots,30$ 
 obtained for the GS expressed in 
 the $\sigma^x$ basis. The second reason, that is more restrictive, come from the 
fact that we do not know the functional dependence on $L$ of the finite-size 
corrections of 
(\ref{Mutual renyi information Ising}). 
These corrections may decay as powers of $\ln L$, that makes the precise evaluation quite difficult using lattice sizes 
$L \lesssim 100$.

It is  interesting  to stress at this point  that all the above results are presumably correct 
if we work in the $\sigma^x$ or $\sigma^z$ basis which correspond to 
free and fixed conformal boundary conditions in the euclidean approach. 
On the other hand we know that in the Ising model we have just these 
two conformal boundary conditions \cite{Cardy1989}.   
Consequently if one works with different basis, other than $\sigma^x$
and $\sigma^z$, one might not get the same results as above because the corresponding boundary conditions are
not conformal. In order to test this we consider the general local basis,

\begin{eqnarray}\label{General base transformation in Q2}
 \begin{bmatrix}
       |a>            \\[0.3em]
       |b>  
     \end{bmatrix}=
 \begin{bmatrix}
      \cos\theta &  \sin\theta e^{-i\alpha}          \\[0.3em]
      \sin\theta e^{-i\phi}          & -\cos\theta e^{-i(\alpha+\phi)}
     \end{bmatrix}
      \begin{bmatrix}
       \ket{\uparrow}            \\[0.3em]
       \ket{\downarrow}   
     \end{bmatrix},
\end{eqnarray}
where $\ket{\uparrow}$ and $\ket{\downarrow}$ are the spin up and down 
components in the $\sigma^z$ basis. We
 calculate the Shannon and R\'enyi entropies in different basis. The numerical results for the $\sigma^y$ basis ($\theta=\pi/4,\alpha=\pi/2,\phi=0$)
 and for another arbitrary $B$ basis where $\theta=\pi/3$, $\alpha=\pi$ and 
$\phi=\pi/5$ 
 are shown in the Fig.~3. We clearly see in this figure 
that the finite-size scaling function 
(\ref{Mutual renyi information Ising}) looks valid even if we chose  non-conformal
basis, however the $n$ dependence of the coefficients are quite different from the  one obtained in the two conformal basis.

\subsection{Mutual information
in the $Q=3$  and $Q=4$ state Potts quantum chain} 
\begin{figure}
\begin{center}
\includegraphics[clip,width=0.9\linewidth]{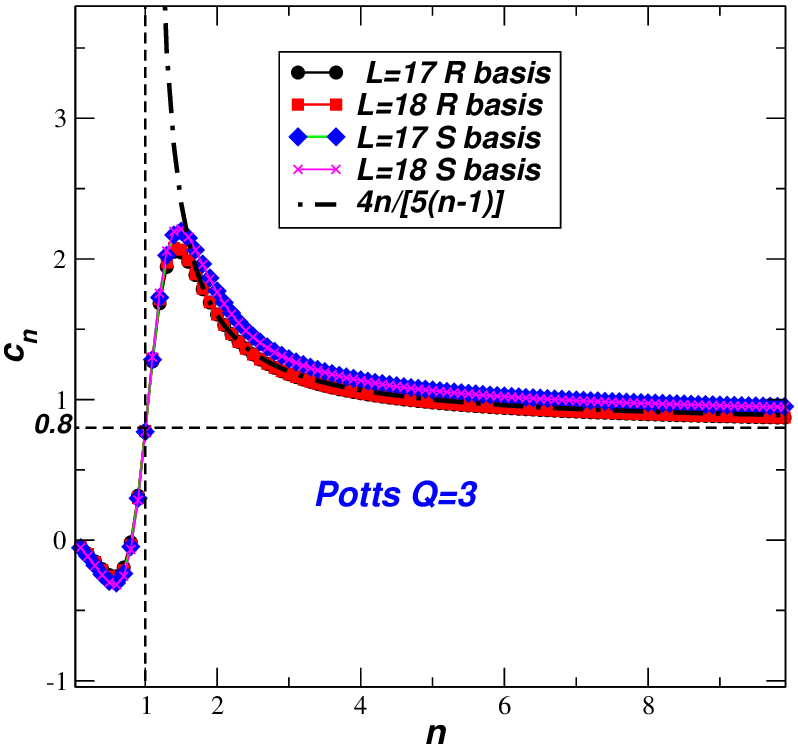}\\
\caption{\label{fig.4} (Color on line) Coefficient of the logarithmic term  of the R\'enyi MI in the $Q=3$ Potts model in the $R$ and $S$ basis \cite{footnote}. 
The coefficients were found by restricting 
the fitting of 
(\ref{Mutual renyi information Ising}) to the subsystem sizes 
 $\ell=4,5,...,{\mbox{Int}}[L/2]$.
The dashed straight lines are guidelines for $n=1$ and for the central charge $c=0.8$.}
\end{center}
\end{figure}
\begin{figure}
\begin{center}
\includegraphics[clip,width=0.9\linewidth]{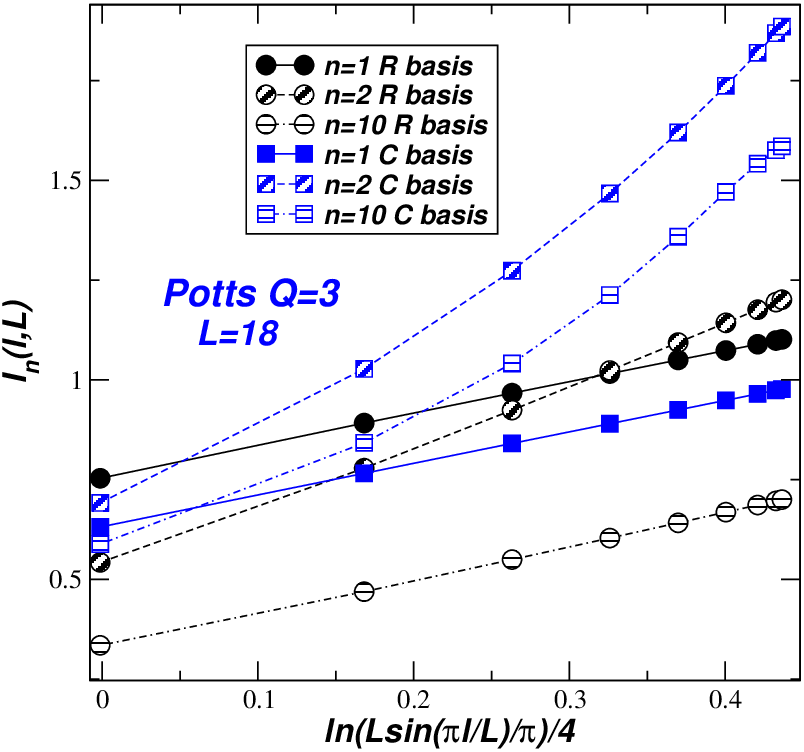}\\
\caption{\label{fig.5} (Color online)
 R\'enyi MI with respect to $\ln \left(\frac{L}{\pi}\sin(\frac{\pi\ell}{L})\right)$ in the $Q=3$ Potts model in the $R$ and $C$ basis 
($\theta,\phi)=(\frac{\pi}{2},\frac{\pi}{4}$). In the $R$ basis the data shows  a good fit for all values  of $n$.  In the $C$
basis (except at $n=1$) the fitting is reasonable only if  we take just the last five or six points. Notice also that, in the large $n$ limit, the linear coefficient of the fitting 
that give $c_n$, are very different in the two basis.}
\end{center}
\end{figure}

The Q-state Potts model in a periodic lattice is defined by the Hamiltonian \cite{Wu1982}

\begin{eqnarray}\label{Potts Hamiltonian}
H_Q=-\sum_{i=1}^L\sum_{k=1}^{Q-1}(S_i^kS_{i+1}^{Q-k}+\lambda R_i^k),
\end{eqnarray}
where $S_i$ and $R_i$ are $Q\times Q$ matrices satisfying the following $Z(Q)$ algebra: 
$[R_i,R_j]=[S_i,S_j]=[S_i,R_j]=0$ for $i\neq j$ and
$S_jR_j=e^{i\frac{2\pi}{Q}}R_jS_j$ and $R_i^Q=S_i^Q=1$.
The system is critical  at the self dual point $\lambda=1$. The critical behavior is governed by a CFT with
 central charge $c=1-\frac{6}{m(m+1)}$ where
$\sqrt{Q}=2\cos(\frac{\pi}{m+1})$. The $Q=2$ Potts chain is just the Ising model which we already discussed 
in the previous section. In this section we will discuss the mutual information of 
the GS in the
$Q=3$ and $Q=4$ Potts chain which follows a similar behavior as that of 
the Ising model. We first summarize our results  regarding 
different basis in the $Q=3$ Potts model. In the basis where the $S$ matrix is diagonal the $S$ and $R$ matrices have the following
forms:
\begin{eqnarray}\label{ S and R matrics}
S = \begin{bmatrix}
       1 & 0 & 0           \\[0.3em]
       0 & \omega          & 0 \\[0.3em]
       0           & 0 & \omega^2
     \end{bmatrix}, \hspace{1cm}
     R = \begin{bmatrix}
       0 & 1 & 0           \\[0.3em]
       0 & 0         & 1\\[0.3em]
       1           & 0 & 0
     \end{bmatrix},
\end{eqnarray}
where $\omega = exp(2\pi i/3)$. 
One can simply get the basis in which  the $R$ matrix is diagonal by just exchanging the two matrices 
 $S\leftrightarrow R$. 

Starting from the $S$ diagonal or from the $R$ diagonal basis 
($\ket{0},\ket{1},\ket{2}$) one can 
introduce  other basis ($\ket{\tilde{0}},\ket{\tilde{1}},\ket{\tilde{2}}$), 
by using the following transformations:
\begin{eqnarray}\label{General base transformation in Q3}
  \begin{bmatrix}
       |\tilde{0}>            \\[0.3em]
       |\tilde{1}>  \\[0.3em]
       |\tilde{2}>  
 \end{bmatrix}= A_3 
  \begin{bmatrix}
       |{0}>            \\[0.3em]
        |{1}>  \\[0.3em]
       |{2}>   
  \end{bmatrix},
\end{eqnarray}
where 
\begin{eqnarray} \label{A3}
 A_3(\theta,\phi)=
\begin{bmatrix}
       \cos\theta & 0 & \sin\theta           \\[0.3em]
     \sin\phi\sin\theta   & \cos\phi          & -\sin\phi\cos\theta \\[0.3em]
       -\sin\theta\cos \phi           & \sin\phi & \cos\theta\cos\phi
     \end{bmatrix},
\end{eqnarray}
is characterized by the angles $\theta$ and $\phi$.
This is not the most general rotation, that depends on the three Euler angles, but is enough for our pouposes. 
 Using this matrix one can express  the 
$S$ and $R$ matrices in a more general basis as
 \begin{eqnarray}\label{S and R matrices in general bases}
\tilde{S}=A_3^{-1}SA_3,\hspace{1cm}\tilde{R}=A_3^{-1}RA_3.
\end{eqnarray}
Having the full structure of the general basis in the 3-state Potts model we calculated the R\'enyi mutual information in 
different basis. As one can see in  Figs.~4 and 5 the $n$-behavior of the R\'enyi  mutual information  depends on 
 the basis that one chooses. For the two basis, $R$ or $S$ diagonal  
(see Fig.~4), this dependence is
\begin{equation} \label{Mutual renyi information Potts}
I_n(\ell,L)= \frac{c_n} {4}\ln \left(\frac{L}{\pi}\sin(\frac{\pi\ell}{L})\right) + ...,
\end{equation}
with

\begin{equation}\label{c-n conformal Potts}
c_n=c\left\{
\begin{array}{c l}      
    1, & n=1\\
\frac{n}{n-1}, & n>1.5 
\end{array}\right. .
\end{equation}
where $c=\frac{4}{5}$ is the central charge of the model. Based on our numerical calculation it is hard to conclude
 the existence or not of 
 a discontinuity at $n=1$, however, if this is the case for the Ising model it is 
likely to be true  also in this model because they follow very similar behavior. Another important point is that
although our results for $n=1$ is consistent with the $c_1=c$ it is very hard to exclude 
the possibility of this number being very close to the central charge and not 
the central charge itself, as claimed in 
\cite{Stephan2014} for the Ising model. 
Note that  (\ref{Mutual renyi information Potts}) is consistent with the picture that $S$ and $R$ basis lead to fixed and free boundary conditions respectively, and 
so can be connected to the bondary CFT as we argued in the case of the Ising 
model.

As one can see in  Fig.~5 the other basis ($C$ basis means that starting from the $S$ basis  we 
choose 
$A_3(\frac{\pi}{2},\frac{\pi}{4})$ in 
(\ref{General base transformation in Q3})) does not follow a similar structure. 
Even if we try
  to fit the data to 
$\ln \left(\frac{L}{\pi}\sin(\frac{\pi\ell}{L})\right)$ by taking just the last four or five points it is clear that the trend for large $n$ is not
compatible with $c_n=c\frac{n}{n-1}$. It is intriguing that even in this basis the results for $n=1$ are quite compatible with the results coming from the conformal basis.
Although we checked few non-trivial basis and not found any other conformal basis our 
study does not necessarily exclude some other possible complicated conformal basis. This is just simply because the
boundary conformal field theory of the 3-state Potts model is much richer than just the two cases (free and fixed) that we 
studied. Finding other possible conformal basis can be very interesting. 
\begin{figure}
\begin{center}
\includegraphics[clip,width=0.9\linewidth]{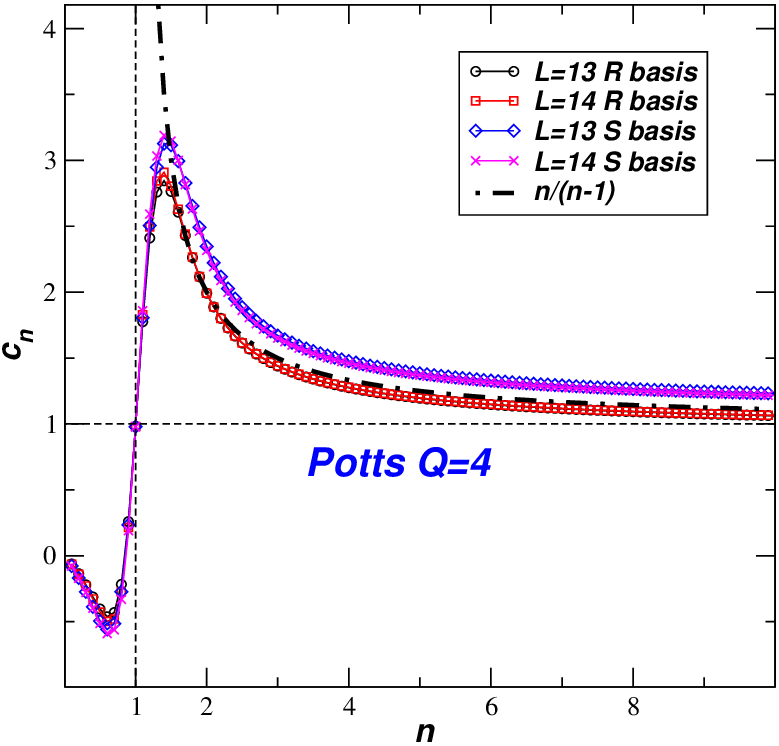}\\
\caption{\label{fig.6} (Color online)
Coefficient of the logarithmic term  of the R\'enyi MI in the $Q=4$ Potts model in the $R$ and $S$ basis \cite{footnote}. The coefficients were found by conditioning
the fitting to the subsystem sizes $\ell=4,5,...,{\mbox{Int}}[L/2]$.
The dashed straight lines are guidelines for $n=1$ and for the central charge $c=1$.}
\end{center}
\end{figure}

We now  study the $Q=4$ Potts model which has a very similar structure as the $Q=3$ Potts model. 
In the basis where the $S$ matrix is diagonal the $S$ and $R$ matrices 
are given by:
\begin{eqnarray}\label{S and R matrics Q3}
S = \begin{bmatrix}
       1 & 0 & 0 &0          \\[0.3em]
       0 & \omega          & 0&0 \\[0.3em]
       0           & 0 & \omega^2&0\\[0.3em]
       0&0&0&\omega^3
     \end{bmatrix}, \hspace{1cm}
     R = \begin{bmatrix}
       0 & 1 & 0 &0          \\[0.3em]
       0 & 0         & 1&0\\[0.3em]
       0&0&0&1\\[0.3em]
       1           & 0 & 0&0
     \end{bmatrix},
\end{eqnarray}
where $\omega=\exp(2\pi i/4)$. 
Like in the $Q=3$ case one can  get the basis which makes the $R$ matrix diagonal by just 
exchanging the two matrices 
 $S\leftrightarrow R$. The most general basis has a complicated form. Here we work with a subset
of the possible non-trivial basis which are obtained  by just using the transformation matrix $A_3$ of the 
$Q=3$ Potts chain. Starting with the basis ($\ket{0},\ket{1},\ket{2},\ket{3}$) 
where $R$ or $S$ is diagonal we obtain the basis  ($\ket{\tilde{0}},
\ket{\tilde{1}},\ket{\tilde{2}},\ket{\tilde{3}}$):

\begin{eqnarray}\label{base transformation in Q4}
 \begin{bmatrix}
       |\tilde{0}>            \\[0.3em]
       |\tilde{1}>  \\[0.3em]
       |\tilde{2}>  \\[0.3em]
       |\tilde{3}>
     \end{bmatrix}=
 \begin{bmatrix}
       \cos\theta & 0 & \sin\theta&0           \\[0.3em]
    \sin\theta\sin\phi    & \cos\phi          & -\sin\phi\cos\theta &0 \\[0.3em]
       -\sin\theta\cos\phi           & \sin\phi & \cos\phi\cos\theta&0 \\[0.3em]
       0&0&0&1
     \end{bmatrix}
      \begin{bmatrix}
       |{0}>            \\[0.3em]
        |{1}>  \\[0.3em]
       |{2}>   \\[0.3em]
       |{3}>
     \end{bmatrix}.
\end{eqnarray}

We have calculated the R\'enyi mutual information in different basis. 
The structure is perfectly compatible with the results for the Ising and $Q=3$ Potts model. The R\'enyi mutual information,
 in the $S$ and $R$ basis, are shown  in Fig.~6. They follow the equations (\ref{Mutual renyi information Potts}) and 
(\ref{c-n conformal Potts}) with $c=1$. 
The difference we see from the results of the two basis is probably due to the 
finite-size corrections since the largest lattice we considered is $L=14$ for the $Q=4$ Potts chain. 
  In the other basis we found a similar structure as we found in the case of the $Q=3$ Potts model (see Fig.~5), indicating that even assuming the  $c_n\ln(\frac{L}{\pi}\sin(\ell\pi/L))$ behavior the coefficient $c_n$ for $n$ large is not given by (\ref{c-n conformal Potts}).
Here we summarize the results for the $Q$-state Potts chain:
\begin{enumerate}
 \item The mutual R\'enyi entropy follows the formulas (\ref{Mutual renyi information Potts}) and 
(\ref{c-n conformal Potts}) in the $S$ and $R$ basis.
  \item In the region $1<n<1.5$ the $c_n$ coefficient has a maximum. Our numerical calculation is consistent
 but non conclusive with the possible presence of discontinuity at $n=1$. 
\item For arbitrary basis the large $n$ behavior of $c_n$ is not given by
(\ref{Mutual renyi information Potts}).
\end{enumerate}

\subsection{Mutual information in the Ashkin-Teller quantum spin chain} 
\begin{figure}
\begin{center}
\includegraphics[clip,width=0.9\linewidth]{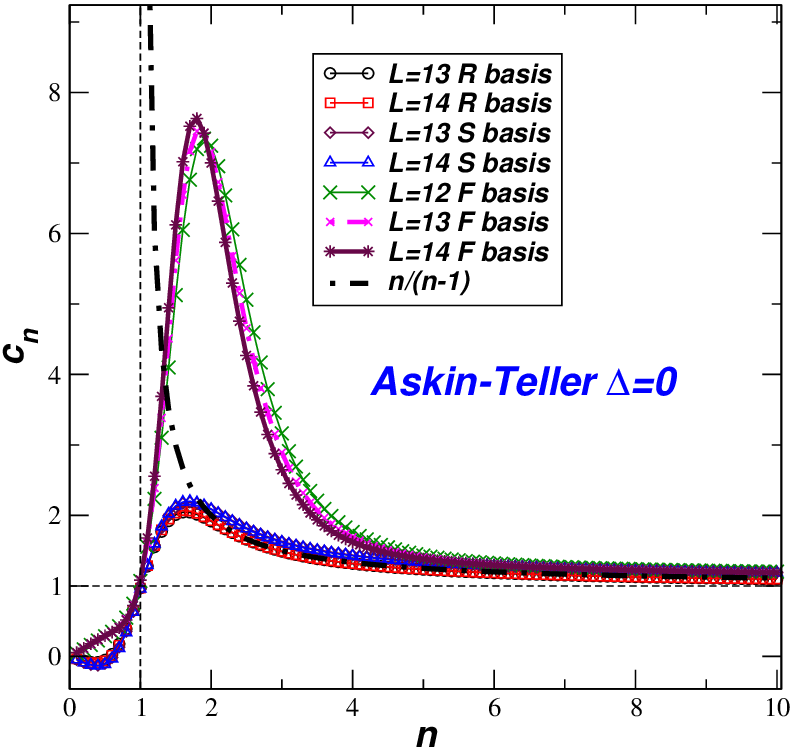}\\
\caption{\label{fig.7} (Color online)
Coefficient of the logarithmic term  of the R\'enyi MI in the Ashkin-Teller model with $\Delta=0$ in the conformal $R$ and $S$ basis and in the $F$ basis 
 specified by the angles $(\theta,\phi)=(\frac{\pi}{4},\frac{\pi}{4})$ in 
(\ref{base transformation in Q4}).
 The coefficients \cite{footnote} were found by conditioning
the fitting to the subsystem sizes $\ell=4,5,...,{\mbox{Int}}[L/2]$.
The dashed straight lines are guidelines for $n=1$ and for the central charge $c=1$.}
\end{center}
\end{figure}
\begin{figure}
\begin{center}
\includegraphics[clip,width=0.9\linewidth]{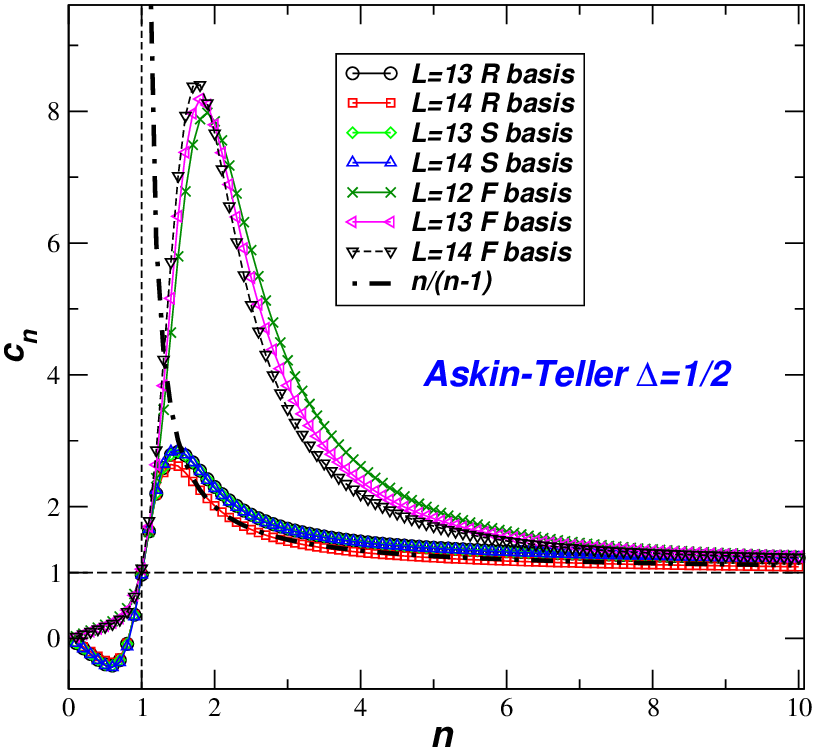}\\
\caption{\label{fig.8} (Color online)
Coefficient of the logarithmic term  of the R\'enyi MI in the Ashkin-Teller model with $\Delta=\frac{1}{2}$ in the conformal $R$ and   $S$ basis and in the $F$ basis 
specified by the angles $(\theta,\phi)=(\frac{\pi}{4},\frac{\pi}{4})$ in 
(\ref{base transformation in Q4}).
 The coefficients \cite{footnote} were found by restricting 
the fitting to the subsystem sizes $\ell=4,5,...,{\mbox {Int}}[L/2]$.
The dashed straight lines are guidelines for $n=1$ and for the central charge $c=1$.}
\end{center}
\end{figure}

The next model that we  study is the Ashkin-Teller model which has a      
$Z(2)\otimes Z(2)$ symmetry  and  whose Hamiltonian is given by:
\begin{equation}
H = -\sum_{i=1}^L\Big{(}[S_i S_{i+1}^3 +S_i^3 S_{i+1} +\Delta S_i^2S_{i+1}^2]  +[R_i +R_i^3 
+\Delta R_i^2]\Big{)},
\end{equation}
where $S$ and $R$  are the same  matrices introduced  in the  
 $Q=4$  Potts model. 
The model is critical and conformal invariant for $-1<\Delta\leq 1$ with the central charge $c=1$. It is worth mentioning
that at $\Delta = 1$ we recover the  $Q=4$ Potts model and at $\Delta =0$ the 
 model  is equivalent to two decoupled Ising models. 
We calculated the R\'enyi mutual information of the GS in different basis for $\Delta=0$ and $\Delta=\frac{1}{2}$. 
The results are shown in the Figs.~7 and 8.  One can summarize the results as follows:
\begin{enumerate}
 \item The mutual Shannon entropy follows the formula 
 \begin{equation}
 I_n(\ell,L-\ell)=\frac{c}{4}\ln \left(\frac{L}{\pi}\sin(\frac{\pi\ell}{L})\right)i + \cdots, \quad c=1,
 \end{equation}
 independent of $\Delta$ in
 the two conformal basis where $S$ and $R$ are diagonal.
 \item The mutual R\'enyi entropy is in general $\Delta$ dependent for $1<n<2$  even in the conformal basis (basis where $S$ or $R$ are diagonal), 
 however, it follows the finite-size scaling function 
 \begin{equation}
 I_n(\ell,L-\ell)=
 \frac{n}{4(n-1)}\ln \left(\frac{L}{\pi}\sin(\frac{\pi\ell}{L})\right)
 \end{equation}
  for $n>2$, independent of the $\Delta$, in the two basis where $S$ or  $R$ are 
diagonal. Presumably as we had  in the $Q=2$ and $Q=3$ cases these two 
basis are also related to the fixed and free conformal boundary conditions.  If we accept the picture that we had in the quantum Potts case one might 
argue that the difference in the two cases $\Delta=0$ and $\Delta=\frac{1}{2}$ in the region $1<n<2$ is just a finite-size effect and, in the limit of large system sizes,  
the results are independent of $\Delta$ in the two conformal basis.
  \item For the non-trivial basis like the $F$ basis, 
obtained by using in 
(\ref{base transformation in Q4})
 $(\theta,\phi)=(\frac{\pi}{4},\frac{\pi}{4})$,   we found that the logarithmic fit is reasonable for both values of 
$\Delta=0,\frac{1}{2}$. 
However the coefficients $c_n$ could be very different from the conformal basis. See Figs.~7 and ~8.
Due to the large and uncontrolled finite-size corrections it is difficult to 
predict a convergence towards the asymptotic behavior $n/(n-1)$.

\end{enumerate}

\subsection{Mutual information in the XXZ quantum spin chain} 

The Hamiltonian of the XXZ chain is defined as 
 \begin{eqnarray}\label{XXZ}
H_{\text{XXZ}}=-\sum_{i=1}^L(\sigma_j^x\sigma_{j+1}^x+\sigma_j^y\sigma_{j+1}^y+\Delta\sigma_j^z\sigma_{j+1}^z),
\end{eqnarray}
where $\sigma^x$,$\sigma^y$ and $\sigma^z$ are spin-$\frac{1}{2}$ Pauli
matrices and $\Delta$ is an anisotropy. 
The model is critical and conformal invariant for
 $-1\leq \Delta < 1$. The long-distance critical fluctuations are 
ruled by a CFT with central charge $c=1$ described by a  compactified boson whose action is given by  
 \begin{eqnarray}\label{compactified Boson}
S=\frac{1}{8\pi}\int d^2x (\bigtriangledown\phi)^2,\hspace{1cm}\phi\equiv\phi+2\pi R,
\end{eqnarray}
where  the compactification radius depends upon the values of $\Delta$, 
namely:
\begin{equation} \label{R}
 R=\sqrt{\frac{2}{\pi}\arccos \Delta}. 
\end{equation}
The Shannon entropy of the system in the $\sigma^z$ basis was already studied in many papers \cite{Hsu2009,Stephan2009,Oshikawa2010}. The analytical and 
numerical results, for the periodic case, indicate that:
 \begin{eqnarray}\label{Shannon XXZ}
Sh(L)=\mu L+ \ln R-\frac{1}{2},
\end{eqnarray}
where $R$ is given by (\ref{R}). The extension of these results to the R\'enyi entropies are  \cite{Stephan2009,Oshikawa2010,Stephan2011}:

\[
 Sh_n(L)= \mu_n L+\begin{dcases*}
       \ln R-\frac{\ln n}{2(n-1)} , & $n<n_c$,\\
        \frac{1}{n-1}(n\ln R -\ln d), &$n\geq n_c$,
        \end{dcases*}
\]
where $n_c=\frac{d^2}{R^2}$  and the parameter $d$ can be understood as the degeneracy of the 
configuration with the highest probability in the ground state. Since in this paper we will always fix the total magnetization
in the $\sigma^z$ basis to zero we will always have $d=2$.

In this section we  extend the above results to the reduced Shannon and the reduced R\'enyi entropies of the 
 quantum chains on their GS.  
An  important point to 
notice is that the techniques used in the previous sub-section for the Ising model are not necessarily applicable in the present  case because the configuration with the 
highest probability in the $\sigma^z$ basis has anti ferromagnetic nature ( for $\Delta \leq 0$) rather than a simple ferromagnetic one\cite{high-prob}. The interesting point is that these kinds of
spin alternating configurations are supposed to be renormalized to Dirichlet boundary conditions in the Luttinger liquid representation of the XXZ model \cite{Affleck} 
and one can hope that they might be connected to the 
underlying CFT \cite{Stephan2013,Stephan2014} ruling the long-distance physics 
of the quantum chain.
We conjecture, see also \cite{Stephan2014}, that 
the reduced R\'enyi entropy for the sub-system size $\ell$, in the $\sigma^z$ basis, is given by
\begin{eqnarray}\label{Renyi periodic BC XXZ}
Sh_n(\ell)=b_n \ell+\frac{c_n}{8}\ln\Big{(}\frac{L}{\pi}\sin(\frac{\pi \ell}{L})\Big{)}+...,
\end{eqnarray}
which consequently leads to the following result for the mutual information 
\begin{equation} \label{Mutual renyi information XXZ}
I_n(\ell,L)= \frac{c_n} {4}\ln \left(\frac{L}{\pi}\sin(\frac{\pi\ell}{L})\right) + ...,
\end{equation}
where $c_n$ is shown in Fig.~9. 
 The  coefficient of the logarithm, in this case is dependent on $n$ and $\Delta$. In an interesting development, in 
 \cite{Stephan2014}, it was conjectured that the form of the $c_n$ follows:
 
\[
 c_n= \begin{dcases*}
       1 , & $n<n_c$,\\
        \frac{n}{n-1}, &$n> n_c$.
        \end{dcases*}
\]
Based on \cite{Stephan2014} at $n=n_c$ the result  has a discontinuity. The presence of the discontinuity at $n=n_c$
is attributed to the  least irrelevant operator $V_d=\cos(\frac{d}{R}\phi)$. As far as $n<n_c$ it was argued in 
\cite{Stephan2014} that this operator is irrelevant and one can get $c_n=1$ by simple Luttinger model arguments. However,
when $n>n_c$ this operator is relevant and consequently the field gets
locked into one of the  minima of the  potential $V_d=\cos(\frac{d}{R}\phi)$. 
This simply leads again to the $\frac{n}{n-1}$ behavior as we had in the Ising model case.
Although our numerical results do not show any 
discontinuity it is consistent with the general arguments in \cite{Stephan2014}. In Fig.~9 one can see the results
of $c_n$ for different values of $\Delta$. Interestingly all of them follows 
the behavior  $\frac{n}{n-1}$ after a value of
$n$  close to $n_c=\frac{4}{R^2}$.
\begin{figure}
\begin{center}
\includegraphics[clip,width=0.9\linewidth]{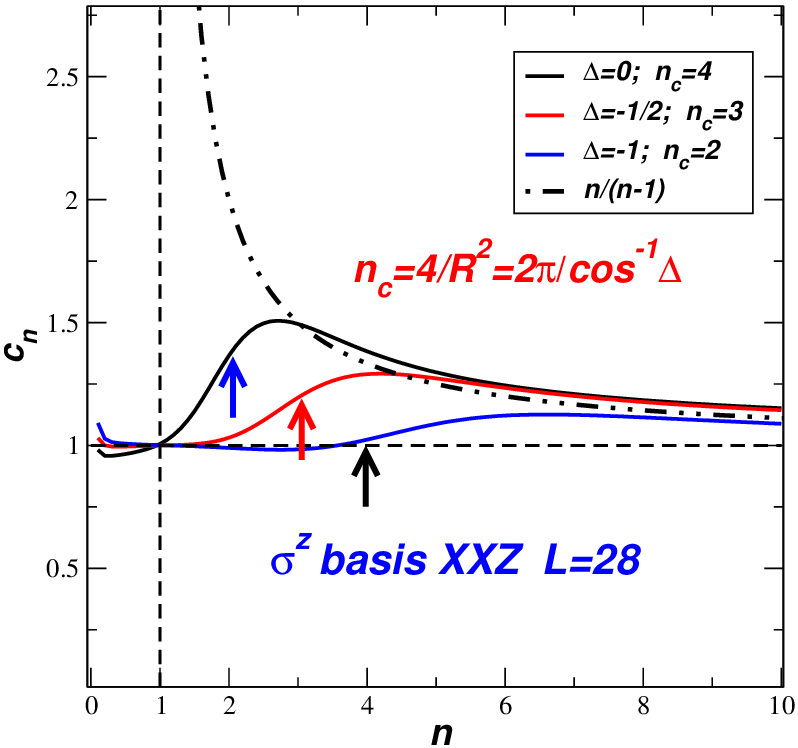}\\
\caption{\label{fig.9} (Color online)
Coefficient of the logarithmic term  of the R\'enyi MI in the XXZ model with 
$L=28$  sites and with different anisotropy parameter $\Delta$ in the $\sigma^z$ basis. 
The coefficients \cite{footnote} were estimated by the average of the fittings 
obtained by restricting the subsystem  sizes to $\ell=4,5,\ldots,14$ and to 
$\ell=5,6,\ldots,14$.
 The arrows  indicate the 
predicted critical value $n_c$, where the asymptotic behavior begins. 
The dashed straight lines are guidelines for $n=1$ and for the central charge $c=1$.}
\end{center}
\end{figure}
\begin{figure}
\begin{center}
\includegraphics[clip,width=0.9\linewidth]{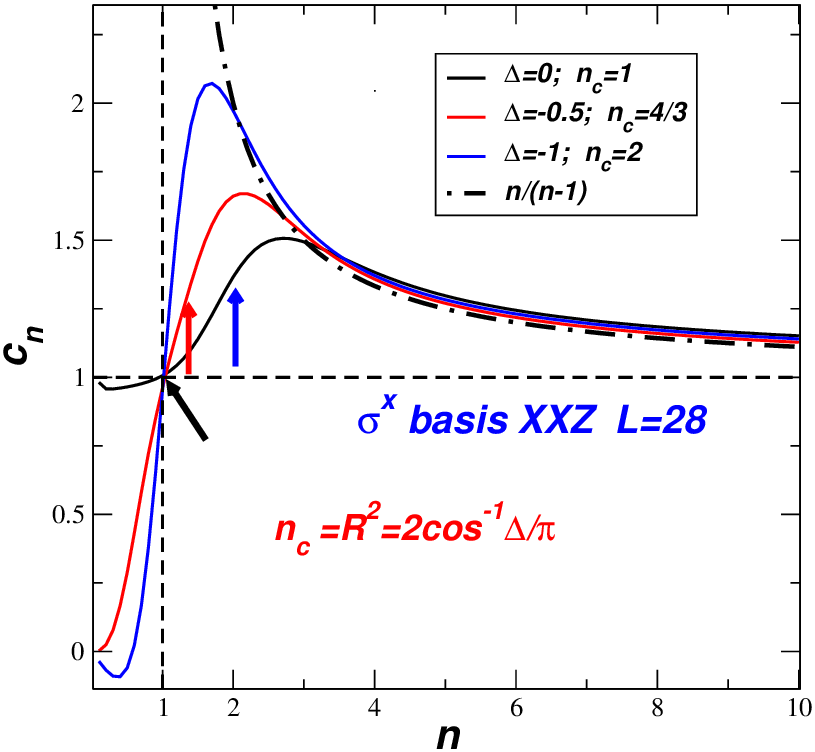}\\
\caption{\label{fig.10} (Color online)
Coefficient of the logarithmic term  of the R\'enyi MI in the XXZ with 
$L=28$ sites and  with different anisotropy parameter $\Delta$ in the $\sigma^x$ basis. 
The coefficients \cite{footnote} were estimated by the average of the fittings 
obtained by restricting the subsystem  sizes to $\ell=4,5,\ldots,14$ and to 
$\ell=5,6,\ldots,14$.
The coefficients were found by restricting 
the fitting to the subsystem sizes $\ell=4,5,...,L/2$. The arrows  indicate  
the predicted  critical value  $n_c$, where the asymptotic behavior begins.
The dashed straight lines are guidelines for $n=1$ and for the central charge $c=1$.}
\end{center}
\end{figure}

One can also do the same kind of analysis  in the other two special basis 
where  $\sigma^x$ or  $\sigma^y$ are diagonal.
Because of the symmetry one expect the same results 
for these two cases and since the basis with fixed $\sigma^x$ is connected to the Dirichlet
boundary condition of the dual field in the Luttinger model representation
\cite{Affleck}  one can simply consider it as the Neumann boundary condition of the Luttinger field. 
This boundary condition is also a conformal boundary condition
and consequently one might hope to be able to find  
the finite-size scaling behavior 
$\frac{c_n}{4}\ln \left(\frac{L}{\pi}\sin(\frac{\pi\ell}{L})\right)$ 
 in the  mutual information calculations. Interestingly one can make the same kind of argument used in the 
 $\sigma^z$ basis and say that the  field $V_d=\cos(dR\tilde{\phi})$, with $\tilde{\phi}\equiv\tilde{\phi}+\frac{2\pi}{R}$
as the dual field, will be relevant
at some value of $n_c=R^2$ and consequently one would expect the 
logarithmic behavior with coefficient $\frac{n}{n-1}$ for $n>n_c$. A very simple check 
for this guess comes from analyzing the point $\Delta=-1$ which is a point which all the basis 
should give the same result because of the $U(1)$ symmetry. Indeed one can simply see that this point has 
$R=\sqrt{2}$ and so both formulas for the critical $n$ give the same answer.

The numerical
results we obtained are consistent  with  the above argument.   The prefactor 
$c_n$  for different $\Delta$'s are shown 
in the Fig.~10. It is  important to stress here that
the results for $n=1$, apart from small deviations that we believe will 
disappear in the $L\rightarrow \infty$,  are  independent of $\Delta$ and equal to the result calculated in the $\sigma^z$ basis. 
However, the results for $n\neq 1$ are in general  different for 
distinct  values of $\Delta$, except when $n>n_c=R^2$, where we found the same 
behavior as we found in the Ising model (or also in the $Q=3$ and $Q=4$ Potts models).   In other words the prefactor of the R\'enyi mutual information
of XXZ model in the $\sigma^x$ basis follows the following formula
\begin{equation}\label{c-n conformal Ising}
c_n=\left\{
\begin{array}{c l}      
    1, & n=1\\
\frac{n}{n-1}, & n>R^2 ,
\end{array}\right.
\end{equation}
Our numerical calculations are not conclusive regarding the presence 
 or absence of a discontinuity in the $c_n$ at $n_c=R^2$. 
 Further numerical calculations with much bigger sizes are needed to make a conclusive argument in this respect.
 In addition based on our numerical results it is not clear that in the regime $1<n<R^2$ 
 the prefactor is constant or not. Another intriguing point is that apart from $\Delta=-1$ case in 
 all the other cases the mutual R\'enyi entropy for $n\to 0$ goes to zero. This behavior is different
 from what we had in the $\sigma^z$ basis.
 \begin{figure}
\begin{center}
\includegraphics[clip,width=0.9\linewidth]{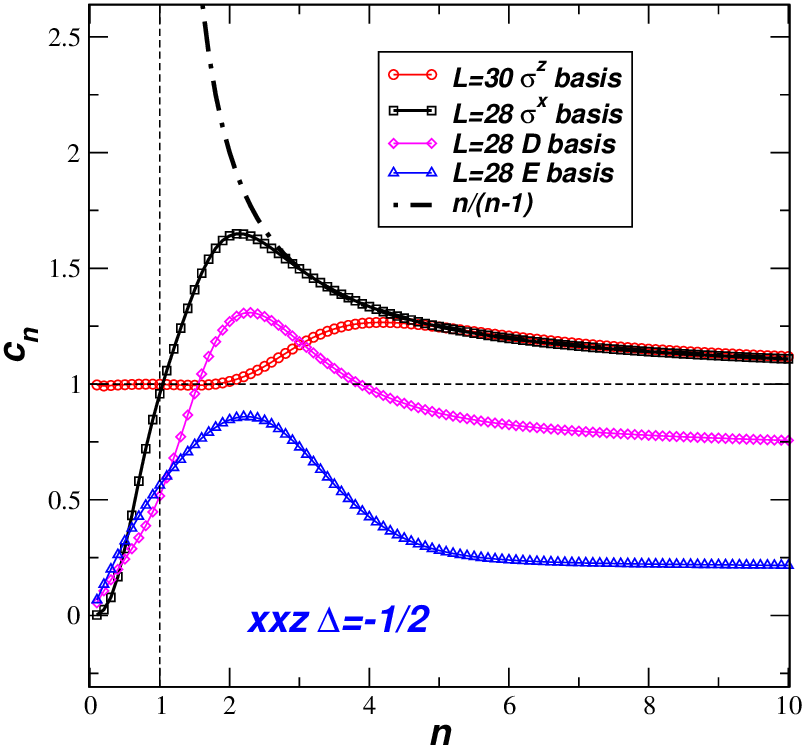}\\
\caption{\label{fig.11} (Color online)
Coefficient of the logarithmic term  of the R\'enyi MI in the XXZ model with $\Delta=-\frac{1}{2}$ in the $\sigma^z$, $\sigma^x$, $D$ and $E$ basis. The 
non-conformal basis $D$ and $E$ are obtained by setting in
 (\ref{General base transformation in Q2}) 
 $(\theta,\pi,\alpha)=(\frac{\pi}{3},\pi,\frac{\pi}{5})$ and  $(\theta,\pi,\alpha)=(\frac{\pi}{2.3},\frac{\pi}{4.5},\frac{\pi}{8.2})$, respectively. 
  The coefficients were found by conditioning the fitting to the 
subsystem sizes $\ell=5,6,...,L/2$.
The dashed straight lines are guidelines for $n=1$ and for the central charge $c=1$.}
\end{center}
\end{figure}
 
 Finally we should stress here that by considering some other basis, i. e.,  
non-conformal basis, will  lead again to  the finite-size scaling function 
 $\frac{c_n}{4}\ln \left(\frac{L}{\pi}\sin(\frac{\pi\ell}{L})\right)$ 
 for the mutual information.
 This is shown for some basis in  Fig.~11. In this figure we 
choose in 
 (\ref{General base transformation in Q2})
the two non-trivial basis $D$ and $E$ where $(\theta,\pi,\alpha)=(\frac{\pi}{3},\pi,\frac{\pi}{3})$ and
 $(\theta,\pi,\alpha)=(\frac{\pi}{2.3},\frac{\pi}{4.5},\frac{\pi}{8.2})$, respectively. However, as we might expected from the results of the previous sections,  the pre factors are not even close to 
  the central charge of the system, differently as happens in the conformal 
basis where $\sigma^z$ or $\sigma^x$ are diagonal.

\section{Conclusions} 

In this paper we have studied different aspects of the mutual Shannon and mutual R\'enyi information  of a bipartite system
 in different quantum critical spin chains such as the Ising model, Q-state Potts model, the Ashkin-Teller model and the 
XXZ quantum chain. We showed that although the MI is in general basis dependent, there are some special basis,  connected with the 
conformal boundary conditions of the underlying CFT, that it is related to  the central charge.  We showed that the general behavior is the same
for the four models: Ising model, $Q=3$ and $4$ Potts models and Ashkin-Teller Model. In all these four
 models the MI calculations, in the conformal basis, show the behavior   $c\frac{n}{4(n-1)}\ln[\frac{L}{\pi}\sin(\pi\ell/L)]$  
 for $n>2$ with a 
possible extension of this regime also to $1<n<2$. At $n=1$ we always get something very close to $\frac{c}{4}$ as the coefficient of the logarithmic term.
 For non-conformal basis the results for the coefficient of the logarithm are completely different and can not be simply related to the central charge of the system. 
In the case of the Ashkin-Teller model we showed that in the conformal basis the results are independent of the anisotropy parameter. We also
studied the same quantities in the XXZ model and showed that in the two conformal basis, where  $\sigma^x$ or $\sigma^z$are diagonal, the results are different. In general one expects
a special value  of $n=n_c$ where beyond this value ($n>n_c$)  the 
finite-size scaling behavior  is 
$c\frac{n}{4(n-1)}\ln[\frac{L}{\pi}\sin(\ell\pi/L)]$.  In more  general basis 
although one can fit the results with a logarithmic function the coefficients do not follow the results obtained in the  conformal basis.

Before closing this paper let us consider again the  possible relationship 
of the Shannon mutual information $I_1(\ell,L)$ with the central charge $c$ of the critical 
chains. In \cite{AR2013}, suggested by the analytical studies of coupled 
harmonic oscillators and by the numerical results of the quantum critical 
chains presented in earlier sections, and also for the spin-1 
Fateev-Zamolodchikov quantum chain, we conjectured that the Shannon mutual
information, like the von Neumann entanglement entropy, is exactly 
related to the central charge of the critical chain: 
$I_1(\ell,L)=\frac{c_1}{4}\ln[\frac{L}{\pi}\sin(\ell\pi/L)] + \gamma_n$, where
$c_1=c$.  The numerical results obtained for all these models, in relative 
small system sizes, deviate from the predicted results, just a few 
percent. In \cite{Stephan2014}, a numerical calculation for the 
quantum Ising model in $\sigma^z$ basis, based on lattice sizes up to $L=56$ indicates that the constant $c_1$ may not be exactly given by the central 
charge but by a close number ($0.480$ instead $0.5$). 
If this disagreement is an effect or not of the unknown finite-size corrections 
is something that only further numerical results with larger lattices can decide.
 This makes the 
problem even more interesting, and rise a natural question: if it is not 
the central charge, what should be this number that is quite close to the 
central charge for quite distinct critical quantum chains? In order to  
further illustrate this problem to other quantum chains we also 
considered the parafermionic $Z_Q$-quantum spin chain \cite{f-z}, with Hamiltonian 
given by \cite{lima,alca}
\begin{equation} \label{zn}
H=-\sum_{i=1}^L\sum_{k=1}^{Q-1} (S_i^kS_{i+1}^{Q-k} + R_i^k)/\sin(\pi k/Q),
\end{equation}
where $S_i$ and $R_i$ are the $Q\times Q$ matrices that appeared in (\ref{Potts Hamiltonian}). This model is critical and conformal invariant with a central charge 
$c=2(Q-1)/(Q+2)$. For the case where $Q=2$ and $Q=3$ we recover the Ising 
and 3-state Potts model, and for case where $Q=4$ we obtain the Ashkin-Teller model with the anisotropy value $\Delta=\frac{\sqrt{2}}{2}$.  
 \begin{figure}
\begin{center}
\includegraphics[clip,width=0.9\linewidth]{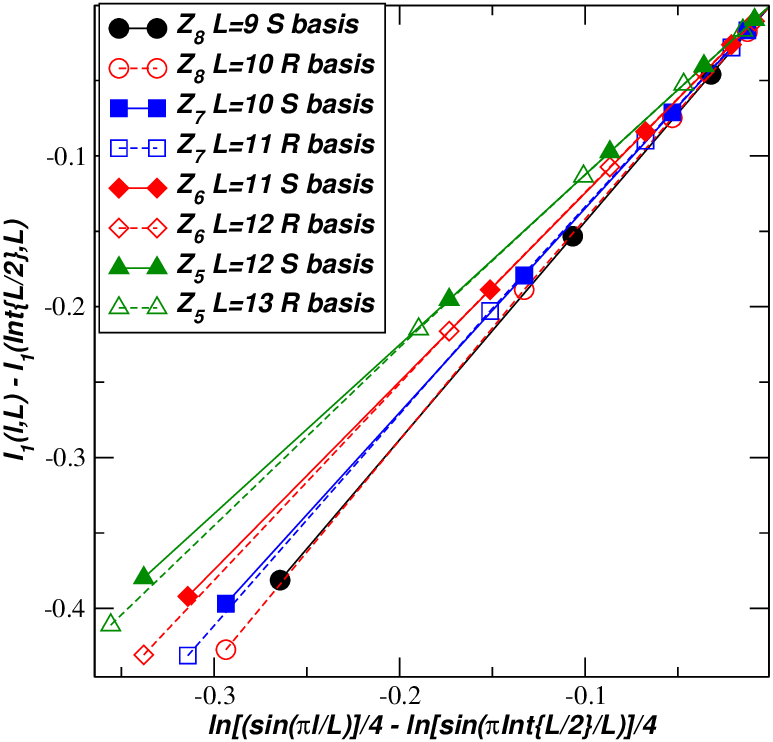}\\
\caption{\label{fig.12} (Color online)
The Shannon mutual information $I_1(\ell,L)$ for the $Z_5$, $Z_6$, $Z_7$, and 
$Z_8$ parafermionic quantum chains with Hamiltonian  given in (\ref{zn}). The results were obtained 
for lattice sizes $L$ and in the basis where $S$ or $R$ is diagonal.}
\end{center}
\end{figure}
\begin{table}[htp]
\caption{Numerical estimates for the constant $c_1$ for the $Z_Q$-parafermionic 
quantum chain  given in (\ref{zn}). The results were obtained using all the  subsystem sizes, with 
the ground-state wavefunction expressed either in $S$ or $R$ basis. The lattice 
sizes used as well the central charge $c=2(Q-1)/(Q+2)$ are also shown.} 
\label{tab1}
\begin{tabular}{lccl}\hline\hline

   $Z_Q$       &basis ($L$) &  $c_1$      &    $c=2(Q-1)/(Q+2)$   \\
\hline
$Z_5$ &  S(12)       &  1.124      & $\frac{8}{7}=1.1427\cdots$   \\
 &  R(13)       &  1.153      &    \\
\hline
$Z_6$ &  S(11)       &  1.250      & $\frac{5}{4}=1.25$   \\
 &  R(12)       &  1.273      &    \\
\hline
$Z_7$ &  S(10)       &  1.352      & $\frac{4}{3}=1.3333\cdots$   \\
 &  R(11)       &  1.372      &    \\
\hline
$Z_8$ &  S(9)       &  1.443      & $\frac{7}{5}=1.4$   \\
 &  R(10)       &  1.456      &    \\
\hline
\hline

\end{tabular}
\end{table}

In Fig.~12 and table~1 we plot the results obtained for the 
$Z_5$, $Z_6$, $Z_7$ and $Z_8$ spin models. We clearly see in Fig.~12 that 
in the basis where either $S$ or $R$ is diagonal,  
except for the first point (subsystem size $\ell=2$), the finite-size scaling 
function is quite well represented by the function $\ln[\sin(\ell \pi/L)]$.
In table~1 we show the results obtained for $c_1$ 
by considering  in the numerical for all the system sizes ($\ell=2,\ldots,{\mbox{Int}}[\frac{L}{2}]$). These results show, like happened in the other models, an estimate of 
$c_1$, for both basis, that deviates a few percent from the central charge. 
It is remarkable that, although the lattice sizes are quite small we 
were able to get values quite close to the predicted central charge. 
We hope that subsequent numerical and analytical studies of the Shannon 
mutual information, that certainly will come,  will shed light to this interesting problem. 
Finally we should emphasize that all the presented results are valid just 
for critical chains. In the gapped phases we expect different behaviors.

\textit{Acknowledgment}:

We would like to thank to P. Calabrese, V. Pasquier, K. Najafi, D. Caravajal Jara, and V. Rittenberg for usefull  
related discussions. This work was supported in part by FAPESP and
CNPq (Brazilian agencies).


\begin{thebibliography}{99}
\bibitem{foot2}
 Distinctly from the von Neumann entanglement entropy of a subsystem of
size $\ell$, that requires  the full tomography of the entire system, the reduced Shannon  entropy of a relatively small subsystem requires only the
tomography of a relative small number of configurations.
\bibitem{high-prob} Although this statement is expected physically we have also 
checked numerically for all lattice sizes we considered.
\bibitem{footnote} The fittings in the region $0<n<0.5$  are 
not as good as for the other values of $n$ since as a function of 
$\ln(\frac{L}{\pi}\sin(\ell\pi/L))$ the data show some curvature as compared 
to the other regions, and the values of $c_n$ are quite small. The negative 
values we found are probably due to finite-size effects.
\end{thebibliography}
\end{document}